\documentstyle[12pt]{article}

\newcommand {\e} {\mbox{\rm e}}

\newcommand {\nn}    {\nonumber}
\newcommand {\vs}[1]  { \vspace*{#1 cm} }

\newcounter{eq}
\newcounter{sc}

\newcommand {\NP}   {Nucl. Phys.}
\newcommand {\PL}   {Phys. Lett.}
\newcommand {\PR}   {Phys. Rev.}
\newcommand {\PRL}   {Phys. Rev. Lett.}

\newcommand {\PTP}  {Prog. Theor. Phys.}

\newcommand {\JMP}  {J. Math. Phys.}

\def\overleftrightarrow#1{\vbox{\ialign{##\crcr
 $\leftrightarrow$\crcr\noalign{\kern-1pt\nointerlineskip}
 $\hfil\displaystyle{#1}\hfil$\crcr}}}

\newlength{\minitwocolumn}
\begin{document}

\begin{flushright}
EDO-EP-40\\
April, 2001\\
\end{flushright}
\vspace{30pt}

\pagestyle{empty}
\baselineskip15pt

\begin{center}
{\large\bf Trapping of Nonabelian Gauge Fields on a Brane

 \vskip 1mm
}

\vspace{20mm}

Ichiro Oda
          \footnote{
          E-mail address:\ ioda@edogawa-u.ac.jp
                  }
\\
\vspace{10mm}
          Edogawa University,
          474 Komaki, Nagareyama City, Chiba 270-0198, JAPAN \\

\end{center}


\vspace{15mm}
\begin{abstract}
We show that as in abelian gauge fields, nonabelian gauge 
fields are also trapped on a brane in the Randall-Sundrum model by 
applying a new mechanism based on topological Higgs mechansim.
It is pointed out that although almost massless gauge fields are
localized on the brane by the new mechanism, exactly massless gauge
fields are not localized. This fact does not yield any problem to
abelian gauge fields, but may give some problem to nonabelian gauge
fields since it is known that there is a discontinuity between massless
and massive fields in the case of the nonabelian gauge theory as in the
gravitational theory.

\vspace{15mm}

\end{abstract}

\newpage
\pagestyle{plain}
\pagenumbering{arabic}

\rm

The brane world scenarios based on the gravity-localized models 
have opened new perspectives for elementary particle physics beyond
the Standard Model \cite{Randall1, Randall2}. (For multi-brane models,
see \cite{Oda0}.) The problems in modern 
physics such as the cosmological constant, the supersymmetry breaking 
and the hierarchy problems might be solved within 
the framework of the brane world scenarios. These problems are hard
since solving such problems seems to require a low energy mechanism,
but not only the low energy physics in the standard framework of effective
theory does not seem to offer a solution but also it is very difficult to 
change the low energy framework in a sensible manner, given all of the 
familiar successes of the Standard Model.

The key ingredient in the gravity-localized models with noncompact
extra spaces is how to localize all the matter and gauge fields in addition 
to the graviton on a brane.
For instance, if charged particles cannot be sharply localized on
the brane, we are in contradiction with the well-established experiment
of the charge conservation law in our world.
Indeed, concerning the localization of various bulk fields on the
brane (or the domain wall), there have been a lot of works thus far
within the framework of local field theory \cite{Jackiw}-\cite{Mouslopoulos}. 
(For a review, see \cite{Oda4}.)
In particular, the non-localization of bulk gauge fields on a brane 
was one of fatal drawbacks associated with the gravity-localized models 
since there certainly exists a massless '$\it{photon}$' in our world.

Recently, we have proposed a new localization mechanism
for abelian gauge fields \cite{Oda5}. 
The idea of constructing such a new mechanism has stemmed from an 
attempt of making a gauge field's analog of the localization mechanism
of bulk fermions. 
The aim of the present paper is to apply the new mechanism to  
nonabelian gauge fields, i.e., '$\it{gluons}$'.
We will see that the nonabelian gauge fields are also localized on the
brane by the same mechanism, but there is a subtle issue associated with 
a discrete difference between the zero-mass theories and the non-zero
mass theories.

Let us first explain the model setup. We consider the following 
$AdS_5$ metric:
\begin{eqnarray}
ds^2 &=& g_{MN} dx^M dx^N  \nn\\
&=& \e^{-A(r)} \eta_{\mu\nu} dx^\mu dx^\nu + dr^2,
\label{1}
\end{eqnarray}
where $M, N, \cdots$ are five-dimensional space-time indices and 
$\mu, \nu, \cdots$ are four-dimensional brane indices. The brane metric 
$\eta_{\mu\nu}$ is the four-dimensional flat Minkowski metric with 
signature $(-,+,+,+)$.  Moreover, $A(r) = 2 k |r|$ where $k$ is a positive 
constant and the fifth dimension $r$ runs from $-\infty$ to $\infty$.
We consider a physical situation such that a single flat 3-brane sits
at the origin of the fifth dimension, $r = 0$, and then ask whether 
nonabelian gauge fields in a bulk can be localized on the brane only by a 
gravitational interaction.  
In this article, we will neglect the back-reaction on the metric from
bulk fields, so we do not need to solve five-dimensional Einstein's 
equations with the energy-momentum tensor of the bulk fields.

Next we shall review the topological massive nonabelian gauge
theories. Such a theory has been first constructed by Yahikozawa and
the present author \cite{Oda6}. In those days, our interests have mainly 
lied in the construction of topological quantum field theories with
the topological Higgs mechanism such that there is no metric tensor field 
in the action and the field equations are merely a flat connection $F = 0$. 
Thus in our theory we have not considered the Yang-Mills action $Tr F^2$, 
for which we need the metric tensors for contraction of space-time indices. 
After our work, the topological massive nonabelian gauge theories with usual 
Yang-Mills kinetic term have been constructed \cite{Hwang, Landim,
Harikumar}. 
In the paper at hand, we shall make use of a theory specified to five 
dimensions of the latter theories. 

In five space-time dimensions, the action of the topological massive 
nonabelian gauge theory is given by \cite{Hwang, Landim, Harikumar}
\begin{eqnarray}
S &=& \int d^5 x \Big[ -\frac{1}{4} \sqrt{-g} g^{M_1 N_1} g^{M_2 N_2} 
F^a_{M_1 M_2} F^a_{N_1 N_2} - \frac{1}{48} \sqrt{-g} g^{M_1 N_1} g^{M_2 N_2} 
g^{M_3 N_3} g^{M_4 N_4}  \nn\\
&\times& {\cal{H}}^a_{M_1 M_2 M_3 M_4} {\cal{H}}^a_{N_1 N_2 N_3 N_4}
+ \frac{m}{12} \varepsilon^{M_1 M_2 M_3 M_4 M_5} C^a_{M_1 M_2 M_3}
F^a_{M_4 M_5} \Big]. 
\label{2}
\end{eqnarray}
Here we have to explain our conventions and notations. Though it is
more convenient to use differential forms in order to express
various formulas, for later convenience we will make use of the conventional 
coordinate-dependent notations. For definiteness, we shall take the gauge
group to be $SU(N)$, with generators $T^a$ satisfying 
\begin{eqnarray}
[T^a, T^b] &=& i f^{abc} T^c, \nn\\
Tr(T^a T^b) &=& \frac{1}{2} \delta^{ab}.
\label{3}
\end{eqnarray}
The field strengths $F$ and $H$ of a gauge field $A$ and a 3-form $C$ are
respectively defined as
\begin{eqnarray}
F^a_{MN} &=& \partial_M A^a_N - \partial_N A^a_M + g f^{abc} 
A^b_M A^c_N, \nn\\
H^a_{MNPQ} &=& 4 (D_{[M} C_{NPQ]})^a \nn\\
&=& (D_M C_{NPQ})^a - (D_N C_{MPQ})^a + (D_P C_{MNQ})^a - 
(D_Q C_{MNP})^a,
\label{4}
\end{eqnarray}
where the covariant derivative $D$ is defined in a usual way as
$(D_M C_{NPQ})^a = \partial_M C^a_{NPQ} + g f^{abc} A^b_M C^c_{NPQ}$.
A newly introduced field strength ${\cal{H}}$ is defined as
${\cal{H}}^a_{MNPQ} = (D_M C_{NPQ})^a + g f^{abc} F^b_{MN} V^c_{PQ}
+$ (cyclic terms) with a 2-form auxiliary field $V$.
(Henceforth, we set the coupling constant $g$ to be 1 except
when the coupling constant is needed.)
This modified field strength ${\cal{H}}$ has been introduced to  
compensate for the non-invariance of the kinetic term, $Tr H^2$, under 
the tensor gauge transformations associated with $C^a_{MNP}$ 
\cite{Thierry}. Note that the $V^a_{MN}$ is a non-dynamical field 
since there is no kinetic term for it in the action.
The action (\ref{2}) is invariant under the gauge transformations
\begin{eqnarray}
\delta A^a_M &=& (D_M \theta)^a = \partial_M \theta^a + f^{abc} 
A^b_M \theta^c, \nn\\
\delta C^a_{MNP} &=& 3 (D_{[M} \omega_{NP]})^a + f^{abc} C^b_{MNP}
\theta^c, \nn\\
\delta V^a_{MN} &=& - \omega^a_{MN} + f^{abc} V^b_{MN} \theta^c.
\label{5}
\end{eqnarray}
The equations of motion following from the action (\ref{2}) read
\begin{eqnarray}
&{}& [ F_{M_1 M_2}, {\cal{H}}^{M_1 M_2 M_3 M_4} ]^a = 0, \nn\\
&{}& \Big( D_{M_2} (\sqrt{-g} F^{M_1 M_2}) \Big)^a 
+ \frac{1}{6} \sqrt{-g} [ C_{M_2 M_3 M_4}, {\cal{H}}^{M_1 M_2 M_3 M_4} ]^a   
\nn\\
&+& \frac{1}{3} \Big( D_{M_2} (\sqrt{-g}
[ V_{M_3 M_4}, {\cal{H}}^{M_1 M_2 M_3 M_4} ]) \Big)^a
- \frac{m}{6} \varepsilon^{M_1 M_2 M_3 M_4 M_5} ( D_{M_2}
C_{M_3 M_4 M_5})^a = 0,                             \nn\\
&{}& \Big( D_{M_4} (\sqrt{-g} {\cal{H}}^{M_1 M_2 M_3 M_4}) \Big)^a 
- \frac{m}{2} \varepsilon^{M_1 M_2 M_3 M_4 M_5} F^a_{M_4 M_5} = 0. 
\label{6}
\end{eqnarray}
It is easy to show that the topological Higgs mechanism in the 
nonabelian case occurs in a set of equations of motion (\ref{6}).

Now using this topological massive nonabelian theory we wish to show that
nonabelian gauge field in a five-dimensional bulk is confined to
a flat brane. For this, we start by taking the gauge conditions of 
the symmetries (\ref{5})
\begin{eqnarray}
A^a_r(x^M) &=& 0, \nn\\
V^a_{MN}(x^M) &=& 0.
\label{7}
\end{eqnarray}
Furthermore, for simplicity we make an ansatz
\begin{eqnarray}
C^a_{rMN}(x^M) = 0.
\label{8}
\end{eqnarray}
With these gauge conditions (\ref{7}) and the ansatz (\ref{8}), we make
the following simple Kaluza-Klein reduction ansatzs for zero-mode 
\begin{eqnarray}
A^a_\mu(x^M) &=& a^a_\mu(x^\lambda) u(r), \nn\\
C^a_{\mu\nu\rho}(x^M) &=& c^a_{\mu\nu\rho} (x^\lambda) u(r).
\label{9}
\end{eqnarray}
where we assume the free equations of motion in four-dimensional flat
space-time:
\begin{eqnarray}
\partial^\mu \bar{f}^a_{\mu\nu} 
= \partial^\mu \bar{h}^a_{\mu\nu\rho\sigma} = 0,
\label{10}
\end{eqnarray}
with the definitions being $\bar{f}^a_{\mu\nu} = 2 \partial_{[\mu} 
a^a_{\nu]}$ and $\bar{h}^a_{\mu\nu\rho\sigma} = 4 \partial_{[\mu} 
c^a_{\nu\rho\sigma]}$.

Even with these simple ansatzs, it is difficult to find the zero-mode
solution $u(r)$ owing to the non-linearity of the equations of motion.
However, as in the analysis of the graviton where the linear fluctuations 
$h_{\mu\nu}$ around the flat metric $\eta_{\mu\nu}$ are studied 
at the lowest level of approximation, it is sufficient to take account 
of the linear equations of the fields, which implies that we look for 
a solution at the lowest level of the coupling constant $g$. The linear 
equations of motion for the fields reduce to 
\begin{eqnarray}
\partial_{M_2} \Big(\sqrt{-g} g^{M_1 N_1} g^{M_2 N_2} 
\bar{F}^a_{N_1 N_2} \Big) - \frac{m}{6} \varepsilon^{M_1 M_2 M_3 M_4 M_5} 
\partial_{M_2} C^a_{M_3 M_4 M_5} &=& 0, \nn\\
\partial_{M_4} (\sqrt{-g} g^{M_1 N_1} g^{M_2 N_2} g^{M_3 N_3} 
g^{M_4 N_4} \bar{H}_{N_1 N_2 N_3 N_4}) 
- m \varepsilon^{M_1 M_2 M_3 M_4 M_5} \partial_{M_4} A^a_{M_5} &=& 0, 
\label{11}
\end{eqnarray}
where $\bar{F}^a_{MN} = 2 \partial_{[M} A^a_{N]}$ and $\bar{H}^a_{MNPQ} 
= 4 \partial_{[M} C^a_{NPQ]}$.
Then, using Eqs. (\ref{1}) and (\ref{9}), Eq. (\ref{11}) takes the forms
\begin{eqnarray}
a^{a\mu} \partial_r \Big( \e^{- A(r)} \partial_r u(r) \Big)
- \frac{m}{6} \varepsilon^{\mu\nu\rho\sigma} c^a_{\nu\rho\sigma}
\partial_r u(r) = 0,
\label{12}
\end{eqnarray}
\begin{eqnarray}
\e^{- A(r)} \partial^\mu a^a_\mu \partial_r u(r)
- \frac{m}{6} \varepsilon^{\mu\nu\rho\sigma} 
\partial_\mu c^a_{\nu\rho\sigma} u(r) = 0,
\label{13}
\end{eqnarray}
\begin{eqnarray}
c^{a\mu\nu\rho} \partial_r \Big( \e^{A(r)} \partial_r u(r) \Big)
- m \varepsilon^{\mu\nu\rho\sigma} a^a_\sigma \partial_r u(r) = 0,
\label{14}
\end{eqnarray}
\begin{eqnarray}
\e^{A(r)} \partial_\rho c^{a\mu\nu\rho} \partial_r u(r)
- m \varepsilon^{\mu\nu\rho\sigma} \partial_\rho a^a_\sigma u(r) = 0.
\label{15}
\end{eqnarray}
Note that up to gauge group index $a$, these equations have the
same forms as those in the case of abelian gauge field \cite{Oda5}.
As the abelian case, since we can regard Eqs. (\ref{13}), (\ref{15})
as the gauge conditions in four-dimensional space-time, the equations 
which we have to solve are Eqs. (\ref{12}), (\ref{14}). 
From the two equations, we can derive a differential equation to $u(r)$:
\begin{eqnarray}
\Big(\partial_r u(r) \Big)^2 - m^2 u^2(r) = 0.
\label{16}
\end{eqnarray}
Provided that after $Z_2$ orbifolding with respect to the fifth dimension 
we impose an even reflection symmetry on $u(r)$, a general solution 
to (\ref{16}) is given by
\begin{eqnarray}
u(r) = c \ \e^{\pm m \varepsilon(r) r},
\label{17}
\end{eqnarray}
where $c$ is an integration constant and $\varepsilon(r)$ is the
step function defined as $\varepsilon(r) = \frac{r}{|r|}$ and 
$\varepsilon(0) = 0$. Moreover, imposing the boundary conditions such that
$u(\pm \infty) = 0, u(0) = u_0$, a special solution takes the form
\begin{eqnarray}
u(r) = u_0 \ \e^{- m \varepsilon(r) r} + {\cal{O}}(g).
\label{18}
\end{eqnarray}
Here in order to emphasize that this solution is a solution up to
the leading order of the gauge coupling constant $g$, we have
put the term ${\cal{O}}(g)$.

We are now ready to check that this zero-mode solution of the gauge
field as well as the 3-form potential leads to a normalizable mode
and the localization on a brane. To do so, plugging Eqs. (\ref{7}),
(\ref{8}) and (\ref{9}) into the starting action (\ref{2}), we have
\begin{eqnarray}
S^{(0)} &=& \int d^4 x \int_{-\infty}^{\infty} dr
\Big[ -\frac{1}{4} F^a_{\mu\nu} F^{a\mu\nu} - \frac{1}{2} 
a^a_\mu a^{a\mu} \e^{-A} (\partial_r u)^2 \nn\\
&-& \frac{1}{48} \e^{2A} H^a_{\mu\nu\rho\sigma} H^{a\mu\nu\rho\sigma} 
- \frac{1}{12} c^a_{\mu\nu\rho} c^{a\mu\nu\rho} \e^{A} 
(\partial_r u)^2 \Big], 
\label{19}
\end{eqnarray}
where the topological term has dropped from the above action 
after performing integration by parts over $r$.  First, let us focus our 
attention to the first plus second terms in (\ref{19}). The calculation 
of the $r$-integrals leads to
\begin{eqnarray}
S_1 &\equiv& \int d^4 x \int_{-\infty}^{\infty} dr
\Big[ -\frac{1}{4} F^a_{\mu\nu} F^{a\mu\nu} - \frac{1}{2} 
a^a_\mu a^{a\mu} \e^{-A} (\partial_r u)^2 \Big]                  \nn\\
&=& \int d^4 x 
\Big[ -\frac{1}{4} \Big\{ \frac{u_0^2}{m} (\partial_\mu a^a_\nu 
- \partial_\nu a^a_\mu )^2 + \frac{4 u_0^3}{3 m} g f^{abc} 
(\partial_\mu a^a_\nu - \partial_\nu a^a_\mu ) a^{b\mu} a^{c\nu} \nn\\
&+& \frac{u_0^4}{2 m} g^2 f^{abc} f^{ade} a^b_\mu a^c_\nu 
a^{d\mu} a^{e\nu} \Big\}
- \frac{1}{2} \frac{m^2 u_0^2}{m + k} a^a_\mu a^{a\mu} \Big], 
\label{20}
\end{eqnarray}
where the coupling constant $g$ was recovered.
In order to transform the free kinetic terms to a canonical form, 
let us redefine the field and the coupling constant as
\begin{eqnarray}
\frac{u_0}{\sqrt{m}} a^a_\mu &\rightarrow& a^a_\mu, \nn\\
\frac{2 \sqrt{m}}{3} g &\rightarrow& g.
\label{21}
\end{eqnarray}
Consequently, the action (\ref{20}) reduces to
\begin{eqnarray}
S_1 &=& \int d^4 x 
\Big[ -\frac{1}{4} \Big\{ (\partial_\mu a^a_\nu - \partial_\nu a^a_\mu )^2 
+ 2 g f^{abc} (\partial_\mu a^a_\nu - \partial_\nu a^a_\mu ) 
a^{b\mu} a^{c\nu} \nn\\
&+& \frac{9}{8} g^2 f^{abc} f^{ade} a^b_\mu a^c_\nu a^{d\mu} a^{e\nu} \Big\}
- \frac{1}{2} \frac{m^3}{m + k} a^a_\mu a^{a\mu} \Big]. 
\label{22}
\end{eqnarray}
Let us note that the coefficient in front of the third term of order
${\cal{O}}(g^2)$ is not 1 but $\frac{9}{8}$, which makes impossible
to transform the parts except a mass term to the square form. This is because
the solution (\ref{18}) is a solution holding only up to ${\cal{O}}(g)$,
thereby meaning that terms of ${\cal{O}}(g^2)$ in the action are ambiguous. 
To determine the correct coefficient in front of the third term, we 
require the gauge invariance for the action except a mass term. 
As a result, the action (\ref{22}) reads 
\begin{eqnarray}
S_1 = \int d^4 x 
\Big[ -\frac{1}{4} f^a_{\mu\nu} f^{a\mu\nu}
- \frac{1}{2} \frac{m^3}{m + k} a^a_\mu a^{a\mu} \Big],
\label{23}
\end{eqnarray}
where $f^a_{\mu\nu}$ is the four-dimensional field strength of
the gauge field $a^a_{\mu}$, $f^a_{\mu\nu} 
= \partial_\mu a^a_\nu - \partial_\nu a^a_\mu + g f^{abc} a^b_\mu a^c_\nu$.
  
Next let turn to the third term in (\ref{19}), that is,
$- \frac{1}{48} \int d^5 x
\e^{2A} H^a_{\mu\nu\rho\sigma} H^{a\mu\nu\rho\sigma}$. Since we
have the expression $H^a_{\mu\nu\rho\sigma} = \partial_\mu 
c^a_{\nu\rho\sigma} u(r) + g f^{abc} a^b_\mu c^c_{\nu\rho\sigma}
u^2(r) +$ (cyclic terms), the $r$-integral in front of the
free kinetic term for $c^a_{\mu\nu\rho}$ becomes
\begin{eqnarray}
I_3 \equiv \int_{-\infty}^{\infty} dr \e^{2A} u^2(r)
= \frac{u_0^2}{m - 2k}.
\label{24}
\end{eqnarray}
when $m - 2k > 0$, whereas it diverges when $m - 2k \le 0$. 
As in abelian gauge field, we assume the relation $m - 2k \le 0$,
by which a 3-form potential $c^a_{\mu\nu\rho}$ is not localized on
a brane, but lives in a bulk away from the brane \cite{Oda5}.
Accordingly, the brane action is given by the action (\ref{23}).
In addition to the relation $m - 2k \le 0$, we have to require the
massless condition of the gauge field, which becomes  
$\frac{m^3}{m + k} \ll 1$. These conditions are simply
satisfied by taking $k \gg m$ as in the case of abelian gauge
field, that is,  '$\it{photon}$'.
Hence we have shown that like abelian gauge field, nonabelian 
gauge field is also localized on a brane by a gravitational interaction.

At this stage, we should comment on one subtlety. Even if we have 
shown that gauge fields with almost vanishing mass can be localized 
on a brane by a gravitational interaction, exactly massless gauge 
fields cannot be localized on the brane by the mechanism mentioned so 
far. This fact can be understood by looking at the normalized zero-mode
in a flat space-time, $\hat{u}(r) = \sqrt{m} \e^{-m \varepsilon(r) r}$,
which becomes vanishing when $m$ is exactly zero. This situation is the 
same as that of fermions where only fermions with mass of a 'kink' profile
can be localized on the brane \cite{Jackiw}.
This problem does not give us any problem to massless gauge fields 
since in the case of the abelian gauge theory the zero-mass case is simply 
the limiting case of the finite mass theory \cite{Veltman}. However, 
it is well-known that in the nonabelian gauge theory and the 
gravitational theory, there is a discrete difference between the theory
with zero-mass and the theory with finite mass, no matter how small as
compared to all external momenta. In particular, this problem is
serious in the gravitational theory since experiments of the bending 
of light rays near the sun and the perihelion movement of Mercury are
distinctly different for the zero-mass graviton and the graviton with
infinitesimally small mass. Of course, experiments force us to select
the zero-mass theory. 

This situation is more subtle for the gauge theory compared to the 
gravitational theory by the following two reasons. In the case of the
gravity, the discontinuity appears at the tree level, while in the
case of the nonabelian gauge theory it occurs at the loop level,
so the corrections might be very small in this case. The second
reason is that although we theoretically regard the gluons as 
exact massless particles, the gluons with a mass as large as a few
MeV may not be precluded experimentally \cite{LBL}. Thus if the
gluons in fact have such a small mass, we can apply the present
localization mechanism to nonabelian gauge fields without 
conflicting experiments. 
 
In conclusion, we have applied a new localization mechanism 
developed for abelian gauge fields to the case of nonabelian
gauge fields.
Our mechanism fully utilizes the topologically massive nonabelian
gauge theories and is very similar to that of fermions in the sense
that in the both mechanisms the zero-modes share the same form and 
the presence of the mass term of a 'kink' profile plays an essential 
role for trapping the zero-modes of the bulk fields on a flat 
Minkowski brane.
It is worthwhile to stress that abelian gauge fields and 
nonabelian gauge fields in a bulk are localized on a brane
by the same mechanism.
We have also pointed out one subtlety associated with massless nonabelian 
gauge fields.

\vs 1


\end{document}